\documentclass[conference,letterpaper]{IEEEtran}
\usepackage[utf8]{inputenc} 
\usepackage[T1]{fontenc}
\usepackage{amsfonts}
\usepackage[cmex10]{amsmath} 
\usepackage{url}
\usepackage{lettrine}
\usepackage{framed}
\usepackage{color}
\usepackage{algorithm}
\usepackage{algorithmic}
\usepackage{overpic}
\usepackage{dsfont}
\usepackage{booktabs} 
\usepackage{framed}
\usepackage{bm}
\usepackage{bbm}
\usepackage{ifthen}
\usepackage{cite}
\usepackage{graphicx}
\usepackage{dsfont}
\usepackage{framed}
\usepackage{bm}
\usepackage{overpic}
\usepackage{xcolor}
\newtheorem{theorem}{Theorem}
\newtheorem{proposition}{Propostion}

\usepackage{xr}

\begin{document}
 
\title{Fundamental Limits of Age-of-Information in Stationary and Non-stationary Environments}

\author{%
  \IEEEauthorblockN{Subhankar Banerjee\IEEEauthorrefmark{1}, Rajarshi Bhattacharjee\IEEEauthorrefmark{2}, Abhishek Sinha\IEEEauthorrefmark{3}}
  \IEEEauthorblockA{Dept. of Electrical Engineering, IIT Madras\\
                    Chennai, India\\ 
                    }
  Email:  
   \IEEEauthorrefmark{1} ee16s048@ee.iitm.ac.in, 
   \IEEEauthorrefmark{2} brajarshi91@gmail.com, 
  \IEEEauthorrefmark{3} abhishek.sinha@ee.iitm.ac.in
                   
}

\maketitle
\begin{abstract}
 We study the multi-user scheduling problem for minimizing the Age of Information (AoI) in cellular wireless networks under stationary and non-stationary regimes. We derive fundamental lower bounds for the scheduling problem and design efficient online policies with provable performance guarantees. In the stationary setting, we consider the AoI optimization problem for a set of mobile users travelling around multiple cells. In this setting, we propose a scheduling policy and show that it is $2$-optimal. Next, we propose a new adversarial channel model for studying the scheduling problem in non-stationary environments. For $N$ users, we show that the competitive ratio of any online scheduling policy in this setting is at least $\Omega(N)$. We then propose an online policy and show that it achieves a competitive ratio of $O(N^2)$. Finally, we introduce a relaxed adversarial model with  channel state estimations for the immediate future. We propose a heuristic model predictive control policy that exploits this feature and compare its performance through numerical simulations.
 \end{abstract}


%

\section{Introduction and Related work}
\lettrine[]{\textbf{T}}{ }he Quality-of-Service (QoS) offered by any wireless network has traditionally been measured along three dimensions, namely, \emph{throughput}, \emph{packet delay}, and \emph{energy efficiency}. There exists an extensive body of literature addressed to optimizing the cross-layer resource allocations to improve the QoS along these axes \cite{tassiulas, mandelbaum2004scheduling, sinha_umw, neely2010stochastic, kozat2004framework}. However, it has been argued that the standard QoS metrics are primarily geared towards quantifying the degree of utilization of the system resources, and less towards measuring the actual user experience \cite{new_QoS}. With the explosive growth of hand-held mobile devices, Internet of Things (IoT), real-time AR and VR systems powered by the emerging 5G technology, the Quality of Experience (QoE) for the users plays a major role in today's network design \cite{QoE}. In order to integrate QoE with the design criteria, a new metric, called \emph{Age-of-Information} (AoI), has been proposed recently for measuring the \emph{freshness} of information available to the end-users \cite{kaul2012real, kosta2017age}.

Designing efficient schedulers to minimize the AoI is currently an active area of research. The papers \cite{kadota2018scheduling} and \cite{kadota2019scheduling} study the \emph{average} AoI minimization problem for static User Equipments (UEs) associated with a single Base Station (BS). In these papers, the authors propose a $4$-optimal Max-Weight-type scheduling policy (Theorem 12 of \cite{kadota2018scheduling}). The paper \cite{srivastava2019minimizing} proposes an optimal scheduling policy for the same setup, where the objective is to minimize the \emph{maximum} AoI of all UEs. All of these papers consider a single-hop network model with static UEs only. The problem of AoI minimization in a multi-hop network with static UEs has been studied in \cite{talak2017minimizing}. The paper \cite{tripathi2019age} considers the problem of designing an AoI-optimal trajectory for a mobile agent which facilitates information dissemination from a central station to a set of ground terminals.  
  The effect of mobility on the capacity of wireless networks has been investigated in the classic work of \cite{grossglauser2002mobility}. It has been shown that mobility, in general, increases the capacity of ad hoc networks. However, to the best of our knowledge, the effect of UE-mobility on the Age-of-Information has not been studied before. One of the main objectives of this paper is to study the AoI-optimal scheduling with mobile UEs. 
   
Most of the existing works on wireless networks assume a stationary channel model for analytical tractability. In rapidly varying environments, such as high-speed trains and vehicle-to-vehicle communication, the standard stationary channel model assumption no longer holds in practice. This is particularly true for the 5G mmWave regime ($\geq 28$ GHz), which suffers from severe attenuation loss \cite{non_stationary, wu2017general}. On the other hand, designing an accurate and analytically tractable non-stationary wireless channel model remains an overarching challenge to the research community \cite{nonstat1, nonstat2}. To overcome this difficulty, in the second part of this paper, we propose a simple adversarial channel model for non-stationary environments and study the scheduling problem in this model. In addition to the emerging 5G technology, the adversarial channel model is also useful for ensuring reliable communication in the presence of tactical jammers, where the interferers, in reality, behave adversarially \cite{poisel2011modern, mpitziopoulos2009survey}. 
\subsection*{Our contributions:}
We make the following contributions in this paper.
\begin{itemize}
\item We study the multi-user scheduling problem in stationary and non-stationary environments. The stationary environment is modelled stochastically, and the non-stationary environment is modelled using an adversarial framework. To the best of our knowledge, this is the first paper that considers the AoI-optimal scheduling problem in an adversarial setting.
\item In the stationary setting described in Section \ref{stochastic}, we design a $2$-optimal scheduling policy for mobile UEs. Our result  improves upon the $4$-optimality bound known for static UEs \cite{kadota2018scheduling, kadota2019scheduling}. 
\item Our analytical result enables us to precisely characterize the effect of mobility on the overall AoI as a function of the long-term user mobility statistics. The results may also be effectively used for small-cell network planning \cite{balazinska2003characterizing}. 
\item In the non-stationary setting of Section \ref{online}, we show that a simple online scheduling policy achieves $O(N^2)$ competitive ratio. Using Yao's minimax principle, we show that no online policy can have a competitive ratio better than $\Omega(N)$.
\item We propose a heuristic scheduling policy in Section \ref{prediction} for the scenario where the future channel states can be accurately estimated for the next $w$ slots. We validate the efficacy of the proposed policy through numerical simulations.
\end{itemize}
The rest of the paper is organized as follows. In Section \ref{sys_model_stochastic}, we describe the stochastic model and formulate the problem in the stationary regime. Section \ref{stochastic} and \ref{online} study the problem in the Stationary and Non-stationary environments respectively. In Section \ref{simulation}, we compare the performance of the proposed scheduling policies via numerical simulations. Section \ref{conclusion} concludes the paper with some pointers to open problems. 

\section{AoI Minimization in Stationary Environments} \label{sys_model_stochastic}
In this section, we first describe the stochastic system model and then formulate the optimal scheduling problem. In the rest of the paper, the abbreviation UE will refer to any generic user equipment, and the term BS will refer to a Base Station. The area covered by a BS will be referred to as a Cell. 
\paragraph{Channel model} We consider a cellular system where a set of $N$ UEs travel around in an area having $M$ BSs. Time is slotted, and at every slot, each BS can beam-form and schedule a packet transmission to one of the UEs in its coverage area. The wireless link to $\textrm{UE}_i$ from the BS in its current cell is assumed to be a stationary erasure channel with the probability of successful reception of a transmitted packet being $p_i, 1\leq i \leq N$.  Hence, when a BS schedules a downlink packet transmission to $\textrm{UE}_i$ in its cell, the packet is either successfully received with probability $p_i$ or lost otherwise. 
\paragraph{Mobility model} We assume that the UE mobility is modelled by a stationary ergodic process. Formally, let the random variable $C_i(t) \in \{1,2,\ldots, M\}$ denote the index of the cell to which $\textrm{UE}_i$ is associated with at time $t$ \footnote{We make the standard assumption that the coverage areas of the cells are mutually disjoint. Hence a UE is associated with only one BS at any time.}. Then, according to our assumption, the stochastic process $\{C_i(t)\}_{t \geq 1}$ is a stationary ergodic process with the probability that $\textrm{UE}_i$ is associated with $\textrm{BS}_j$ at any time $t$ given by $\mathbb{P}(C_i(t)=j)= \psi_{ij}, \forall i,j, t.$ The probability measure $\bm{\psi}$ denotes the stationary occupancy distribution of the cells by the UEs. The mobility of different UEs is assumed to be independent of each other. Many different mobility models proposed in the literature fall under the above general scheme, including the i.i.d. mobility model, random walk model, and the random waypoint model \cite{ge2016user, akyildiz2000new, johnson1996dynamic, bai2004survey}. See Figure \ref{AoI_mobility_fig} in the Appendix \ref{lb_proof} for a schematic. 

\paragraph{Packet arrival model to BS} We consider a \emph{saturated} traffic model, where at the beginning of any slot, each BS receives a fresh update packet from a common external source (\emph{e.g.,} a high-speed optical backbone network). Since the UEs are interested in the latest updates only, the BS then deletes any old packet from its buffer and schedules the fresh packet for transmission to some UE following a scheduling policy. The saturated traffic model is standard in applications relying on continuous status updates \cite{costa2016age}, such as monitoring and surveillance with sensor networks \cite{javani2019age}, velocity and position updates for autonomous vehicles \cite{kaul2011minimizing}, command and control information exchange in mission-critical systems, disseminating stock-index updates and live game scores.

\paragraph{System states} 
For slot $t$, let $t_i(t) < t$ denote the last time before time $t$ at which $\textrm{UE}_i$ received a packet successfully from any BS. The Age-of-Information $h_i(t)$ of $\textrm{UE}_i$ at time $t$ is defined as 
\[ h_i(t) \equiv t-t_i(t). \]
In other words, the random variable $h_i(t)$ denotes the length of time elapsed since $\textrm{UE}_i$ received its last update before time $t$. Hence, the r.v. $h_i(t)$ quantifies the \emph{staleness} of information available to $\textrm{UE}_i$. See Figure \ref{AoI_fig} in the Appendix for a typical evolution of $h_i(t)$. The state of the UEs at time $t$ is completely specified by the Age-of-Information of all UEs, given by the random vector $\bm{h}(t)\equiv \big(h_1(t), h_2(t), \ldots, h_N(t)\big)$, and the association of the UEs with the cells, represented by the cell-occupancy vector $\bm{C}(t)$.  

\paragraph{Policy space and performance metric}
 A scheduling policy $\pi$ first selects a UE in each cell (if the cell contains any UE), and then schedules the transmission of the latest packet from the BSs to the UEs over the wireless erasure channel described earlier. The scheduling decisions are required to be causal for it to be implementable in real-time. The set of all admissible scheduling policies is denoted by $\Pi$. 
Our goal in this paper is to design a distributed scheduling policy which minimizes the long-term average AoI of all users. In view of this, we consider the following average-cost problem:
\begin{eqnarray} \label{objective}
\textsf{AoI}^*=\inf_{\bm{\pi} \in \Pi }	\limsup_{T \to \infty} \frac{1}{T} \sum_{t=1}^{T}\frac{1}{N}\bigg(\sum_{i=1}^{N} \mathbb{E}^\pi(h_i(t))\bigg).
\end{eqnarray}

\subsection{Converse and Achievability} \label{stochastic}
The AoI minimization problem given by \eqref{objective} is an example of an average-cost MDP with countably infinite state-space \cite{bertsekas1995dynamic}. Excepting a few cases with special structures (\emph{cf.} \cite{srivastava2019minimizing}), such problems are notoriously difficult to solve exactly. Moreover, the standard numerical approximation schemes for infinite-state MDPs typically do not provide theoretical performance guarantees. In this paper, we take a different approach to approximately solve the problem \eqref{objective}. In the following Theorem, we obtain a fundamental lower bound to the optimal AoI. Finally, in Theorem \ref{achievability_thm}, we show that a simple online scheduling policy $\pi^{\textsf{MMW}}$ achieves the lower bound within a factor of $2$. 
\begin{framed} 
\begin{theorem}[Converse] \label{lb}
In the stationary setup, the optimal \textsf{AoI} in \eqref{objective} is lower bounded as:
 \begin{eqnarray} \label{lb_expr}
 \textsf{AoI}^* \geq 	\frac{1}{2N g(\bm{\psi})}	\bigg(\sum_{i=1}^{N} \sqrt{\frac{1}{p_i}}\bigg)^2+ \frac{1}{2},
 \end{eqnarray}
where the quantity $g(\bm{\psi})$ denotes the expected number of cells with \emph{at least one UE}, where the expectation is taken with respect to the stationary occupancy distribution $\bm{\psi}$. In particular, since $g(\bm \psi) \leq  \min \{M, N\},$ we also have the following (loose) lower bound which is agnostic of the UE mobility statistics:
 \begin{eqnarray*}
 \textsf{AoI}^* \geq 	\frac{1}{2N \min \{M, N\}}	\bigg(\sum_{i=1}^N \sqrt{\frac{1}{p_i}}\bigg)^2 + \frac{1}{2}.
 \end{eqnarray*} 
\end{theorem}
\end{framed}
Please refer to Appendix \ref{lb_proof} for a proof of this theorem.
\paragraph*{Discussion} Theorem \ref{lb} gives a universal lower bound for the minimum AoI achievable by \emph{any} admissible scheduling policy $\pi \in \Pi$. Interestingly, it reveals that the lower bound depends on the mobility of the UEs only through their stationary cell-occupancy distribution $\bm{\psi}$. Hence, given the stationary distribution $\bm \psi$, the lower bound \eqref{lb_expr} is agnostic of the details of the mobility model. The appearance of the quantity $g(\bm \psi)$ in the lower bound should not be surprising as it denotes the \emph{typical} number of non-empty cells at a slot in the long run. Since a BS can transmit a packet only if at least one UE is present in its coverage area, the quantity $g(\bm \psi)$, in some sense, represents the \emph{multi-user diversity} of the system.

\subsubsection*{Expression for $g(\bm \psi)$}
To get a sense of the lower bound \eqref{lb_expr}, we now work out a closed-form expression for $g(\bm \psi)$ for the uniform UE mobility pattern. 
Using linearity of expectation, 
\begin{eqnarray}\label{g_psi_eq}
	g(\bm{\psi})&=& \mathbb{E}_{\bm \psi} \sum_{j=1}^{M} \mathds{1}(\textrm{BS}_j \textrm{ contains at least one UE}\big)\nonumber \\ 
	&=&\sum_{j=1}^{M} \mathbb{P}_{\bm \psi} \big( \textrm{BS}_j \textrm{ contains at least one UE}\big). 
\end{eqnarray}
Since the cells are disjoint, we readily conclude from \eqref{g_psi_eq} that $g(\bm{\psi}) \leq \min\{M, N\}$. Recall that $\psi_{ij}$ denotes the marginal probability that the $\textrm{UE}_i$ is in $\textrm{BS}_j$. Since the mobility of the UEs are independent of each other, the expected number of non-empty cells $g(\bm \psi)$ in Eqn.\ \eqref{g_psi_eq} simplifies to:
\begin{eqnarray} \label{g_eqn3}
g(\bm{\psi})= \sum_{j=1}^{M} \big(1-\prod_{i=1}^N(1-\psi_{ij})\big).	
\end{eqnarray}
%
%
%
%
We now evaluate the above expression for the 
 case when the limiting occupancy distribution of each UE is \emph{uniform} across all BSs, \emph{i.e.,} $\psi_{ij}=\frac{1}{M}, \forall i,j $. The uniform stationary distribution arises, for example, when the UE mobility can be modelled as a random walk on a regular graph \cite{lovasz1993random}. In this case, Eqn.\ \eqref{g_eqn3} simplifies to
 \begin{eqnarray} \label{g_spl}
g(\bm{\psi^{\textsf{unif}}}) = M \bigg(1-\big(1-\frac{1}{M}\big)^N\bigg).  	
\end{eqnarray}
For $M=1$, we have $g(\bm{\psi})=1$. For $M\geq 2$, we have the following bounds which are easier to work with
 \begin{eqnarray} \label{g_unif}
 M\bigg(1-e^{-\frac{N}{M}}\bigg)\leq g(\bm{\psi^{\textsf{unif}}}) \leq M\bigg(1-e^{-1.387 \frac{N}{M}}\bigg).	
 \end{eqnarray}
For a derivation of the bounds in \eqref{g_unif}, please refer to Appendix \ref{g_unif_proof}.

\subsubsection*{Achievability} \label{achievability}
We now propose an online scheduling policy $\pi^{\textsf{MMW}}$ which approximately minimizes the average AoI \eqref{objective} for mobile UEs (the abbreviation \textsf{MMW} stands for ``Multi-cell Max-Weight"). Our policy is a multi-cell generalization of the $4$-approximate single-BS scheduling policy proposed in \cite{kadota2018scheduling}. Moreover, using a tighter analysis, we give an improved $2$-factor approximation guarantee for $\pi^{\textsf{MMW}}$. 
\paragraph*{The policy $\pi^{\textsf{MMW}}$} 
At every slot, each BS schedules a UE under its coverage that has the highest index among all other UEs. The index $I_i(t)$ of $\textrm{UE}_i$ is defined as $I_i(t) \equiv p_ih_i^2(t).$
\begin{framed}
\begin{theorem}[Achievability]\label{achievability_thm}
	$\pi^{\textsf{MMW}}$ is a $2$-approximation scheduling policy for statistically identical UEs with i.i.d. uniform mobility (\emph{i.e.,} $p_i=p,  \forall i$ and $\psi_{ij}=\frac{1}{M}, \forall i,j$).
\end{theorem}
\end{framed}
For a proof of Theorem \ref{achievability_thm}, please refer to Appendix \ref{achievability_thm_proof}. When the BSs employ power-control, all UEs experience the same SINR, and they become statistically identical. It can be easily seen that the policy $\pi^{\textsf{MMW}}$ is fully distributed and may be implemented with local information only. 
\subsubsection*{Effect of mobility on AoI}
Recall that, a BS can schedule a transmission to only one UE in its cell at every slot. Hence, if  all of the $N$ UEs remain stationary at a single cell, they all have to contend with each other for scheduling. This naturally increases the average AoI of the UEs. On the other hand, if the UEs are mobile, they can take advantage of multiple downlink transmission opportunities from multiple BSs. This form of \emph{multi-user diversity} drastically reduces the overall AoI, by improving the network resource utilization. Next, we quantify the effect of mobility on the average AoI.\\
Define the \emph{Mobility Advantage on AoI} ($\alpha$) to be the ratio of the optimal AoI when all UEs are stationary at a single BS (\emph{i.e.}, $M=1$.) vs. the optimal AoI when the UEs are mobile. As noted above, for a single BS, we have $g(\bm \psi)=1.$
From our achievability result in Theorem \ref{achievability_thm}, we know that the lower bound in Eqn. \eqref{lb_expr} is achievable within a factor of $2$. This implies that $\alpha = \Theta(g(\bm{\psi})).$
From the equation \eqref{g_unif}, we have 
\begin{eqnarray} \label{mobility_advantage}
	g(\bm{\psi^{\textsf{unif}}}) = M\bigg( 1 - e^{-c \frac{N}{M}}\bigg),
\end{eqnarray}
for some constant $1 \leq c \leq 1.387$. Consider the following three scaling regime:
\begin{itemize}
\item \textbf{Constant Density:} If $N$ and $M$ scale in such a way that the \emph{density} of the UEs remains constant, \emph{i.e.}, $\frac{N}{M}=\rho, $ we see that the average AoI diminishes linearly with the number of BSs, \emph{i.e.,} $\alpha = M(1-\exp(-c\rho))$.   	
\item \textbf{Under-Loaded BS:} If $N/M << 1$, we have 
$\alpha \approx M\big(1-1+c\frac{N}{M}\big) = \Theta(N).$	
\item \textbf{Over-Loaded BS:}  If $N/M >>1 $, we have  $\alpha = \Theta(M)$.
\end{itemize}

\section{AoI Minimization in Non-Stationary Environments} \label{online} 
In this Section, we consider the problem of AoI-optimal scheduling with $N$ static users in a non-stationary environment. Since non-stationary channels are difficult to model and analyze, we propose a new adversarial channel model in this setting. Besides being analytically tractable, all positive results in this model (\emph{e.g.,} Theorem \ref{comp_ratio_ub}) carry over to less adversarial environments.
\paragraph*{Channel model} A set of $N$ UEs are under the coverage of a single BS (\emph{i.e.,} $M=1$). The BS can transmit to any one UE at a slot. The channel state $\textsf{Ch}_i(t)$ of any $\textrm{UE}_i$ at any time slot $t$ could be either \textsf{Good} ($1$) or \textsf{Bad} ($0$).  If the BS schedules a packet to a UE having a \textsf{Good} channel at that slot, the UE decodes the packet successfully. Otherwise, the packet is lost. We assume that, the states of the $N$ channels (corresponding to $N$ different UEs) are selected by an \emph{omniscient adversary} from the set of all possible $2^N$ states at every slot. The scheduling policy is \emph{online} and has no information on the channel states for the current or future slots. We will partially relax this assumption in Section \ref{prediction}, by considering a more general class of adversarial channel models with future channel estimations. The cost function over a horizon of $T$ slots is given by:
\begin{eqnarray}\label{cost_fn}
\textsf{AoI}(T) = \sum_{t=1}^{T}\bigg(\sum_{i=1}^N h_i(t)\bigg). 
\end{eqnarray}
The packet arrival model to the BS remains the same as in the stationary environment in Section \ref{sys_model_stochastic}. 
\paragraph*{Performance Metric}
As standard in the literature on online algorithms \cite{fiat1998online, albers1996competitive}, we gauge the performance of an online scheduling policy $\mathcal A$ using \emph{competitive ratio} ($\eta^{\mathcal A}$), which compares the cost of $\mathcal A$ with that of an optimal \emph{offline} policy \textsf{OPT} equipped with hindsight knowledge. More precisely, let $\bm{\sigma} \in \{\{0,1\}^N\}^T$ be a sequence of length $T$ representing the vector of channel states chosen by the adversary for the entire horizon. Then, the competitive ratio of the policy $\mathcal A$ is defined as \cite{albers1996competitive}:
\begin{eqnarray}\label{comp_rat_def}
\eta^{\mathcal{A}} = \sup_{\bm \sigma}\bigg(\frac{\textrm{Cost of the online policy } \mathcal A \textrm{ on } \bm{\sigma}}{\textrm{Cost of OPT on } \bm{\sigma}}\bigg),	
\end{eqnarray}
 where the supremum is taken over all finite-length input sequences $\bm \sigma$, and the cost function is given by \eqref{cost_fn}. In the definition 
 \eqref{comp_rat_def}, while the online policy $\mathcal A$ has only  causal information, the policy \textsf{OPT} is assumed to be equipped with full knowledge on the entire channel-state sequence $\bm \sigma.$ 
\subsection*{Characterization of the optimal offline (\textsf{OPT}) policy}
For a given sequence of channel states $\bm \sigma$ of length $T$, the optimal offline policy \textsf{OPT} may be obtained by using Dynamic Programming. 
Let the variable $C_t^*(h_1(t), h_2(t), \ldots, h_N(t))$ denote the optimal cost-to-go from time $t$ when the AoIs of the the $N$ UEs are given by the vector $\bm{h}(t)\equiv (h_1(t), h_2(t), \ldots, h_N(t)).$ Using standard notations, we have the following backward DP recursion 
\begin{eqnarray}\label{opt_dp}
C^*_{t}(\bm{h}(t))&=& \underbrace{\sum_{i=1}^N h_i(t)}_{\textrm{cost for slot } t} + \underbrace{\min_{i: \textsf{Ch}_i(t+1)=1} C^*_{t+1}(\bm{h}_{-i}(t)+\bm 1, 1)}_{\textrm{optimal future cost}}, \nonumber \\
C^*_{T+1}(\bm{h})&=&0 \hspace{10pt} \forall \bm{h},
\end{eqnarray}
 where the minimization in Eqn.\ \eqref{opt_dp} is over all UEs $i$ having a \textsf{Good} channel at slot $t+1$. When there is no UE with a \textsf{Good} channel at slot $t+1$ (\emph{i.e.,} $\textsf{Ch}_i(t+1)=0, \forall i$), the second term denoting the future cost is replaced with $C^*_{t+1}(\bm{h}(t)+\bm{1})$.  
 \paragraph*{Comparison with the throughput maximization problem} It is interesting to note that the competitive ratio for the sum-throughput maximization problem in this adversarial model can be arbitrarily bad (\emph{i.e.}, unbounded). It can be understood from the following. Consider a system with two users. If an online scheduler $\mathcal{A}$ schedules $\textrm{UE}_1$ at any slot, the adversary can set the channel corresponding to $\textrm{UE}_1$ to \textsf{Bad} and set $\textrm{UE}_2$'s channel to \textsf{Good} and vice versa. At any slot, the optimal policy schedules the user with the \textsf{Good} channel state. Hence, any online scheduler $\mathcal{A}$ receives zero throughput, but \textsf{OPT} achieves the full throughput of unity.   \\ 
 Surprisingly enough, Theorem \ref{comp_ratio_ub} shows that the \textsf{Max Age} (MA) scheduling policy, which schedules a user having the \emph{highest age} (\emph{i.e.,} Scheduled UE at time $t$ $\in \arg\max_i h_i(t)$), is $O(N^2)$-competitive for minimizing the AoI. 
\begin{framed}
\begin{theorem}[Achievability] \label{comp_ratio_ub}
	In the adversarial setting with $N$ users, the \textsf{MA} policy is $O(N^2)$ competitive for minimizing the average AoI.  
\end{theorem}	
\end{framed}
For a proof of Theorem \ref{comp_ratio_ub}, please refer to Appendix \ref{comp_ratio_ub_proof}. On a related note, in our recent work \cite{srivastava2019minimizing}, we showed that the \textsf{MA} policy is exactly optimal for minimizing the \emph{maximum} AoI of all UEs in the stochastic setting. 
\subsection{A Lower bound to the competitive ratio} \label{competitive_ratio_lb}
In this section, we use Yao's minimax principle for obtaining a universal  lower bound to the competitive ratio \eqref{comp_rat_def} in the adversarial setting. In connection with online problems, Yao's minimax principle may be stated as follows:
\begin{framed} 
\begin{theorem}[Yao's Minimax principle \cite{albers1996competitive}]
	Given any online problem, the competitive ratio of the best randomized online algorithm against any oblivious adversary is equal to the competitive ratio of the best deterministic online algorithm under a worst-case input distribution. 
	\end{theorem}
\end{framed}
Using the above principle, it is clear that a lower bound to the competitive ratio of \emph{all} deterministic online algorithms under \emph{any} input channel state distribution $\bm p$ yields a lower bound to the competitive ratio in the adversarial setting, \emph{i.e.,} 
\begin{eqnarray}\label{Yao_lb}
\eta \geq \frac{\mathbb{E}_{\bm{\sigma} \sim \bm{p}}(\textrm{Cost of the Best Deterministic Online Policy})}{\mathbb{E}_{\bm \sigma \sim \bm p}\textrm{(Cost of OPT)}}.	
\end{eqnarray}
To apply Yao's principle in our setting, we construct the following distribution $\bm{p}$ of the channel states:  at every slot $t$, a UE is chosen independently and uniformly at random, and assigned a \textsf{Good} channel. The rest of the UEs are assigned \textsf{Bad} channels.
The rationale behind the above choice of the channel state distributions will become clear when we compute \textsf{OPT}'s expected cost in Appendix \ref{comp_ratio_lb_proof}. In general, the cost of the optimal offline policy is obtained by solving the Dynamic Program \eqref{opt_dp}, which is difficult to analyze. However, with our chosen channel distribution $\bm{p}$, we see that only one UE's channel is in \textsf{Good} state at any slot. This greatly simplifies the evaluation of \textsf{OPT}'s expected cost.  The following Theorem gives the universal lower bound:
\begin{framed}
\begin{theorem}[Converse] \label{comp_ratio_lb}
	In the adversarial set up, the competitive ratio $\eta$ of any online policy with $N$ UEs is lower bounded by $\frac{N}{2}+ \frac{1}{2N}.$ Further, for  $N=2$ UEs, the lower bound can be improved to $1.5.$
\end{theorem}	
\end{framed}
Please refer to Appendix \ref{comp_ratio_lb_proof} for a proof of this Theorem.
 

\subsection{AoI minimization with Channel Predictions} \label{prediction}
The converse result in Theorem \ref{comp_ratio_lb} states that under the adversarial channel model, \emph{any} online scheduling policy has a worst-case competitive ratio $\eta$ which grows at least linearly with the number of UEs ($N$). This is quite a disappointing result when the number of UEs is large. On the flip side, the fully adversarial channel model may also be too restrictive in practice. To circumvent this situation, we now exploit the physical fact that wireless channels with block-fading may often be estimated quite accurately for a few subsequent future slots \cite{prediction_RHC}. We consider a relaxed adversarial model, where at any slot $t$, the BS can estimate the channels perfectly for a window of the next $w \geq 0$ slots. Here, $w$ is an adjustable system parameter that can be adaptively tuned by the policy in accordance with the scale of time-variation of the channels (\emph{e.g.,} fading block length). 
Similar to the adversarial model in Section \ref{online}, we continue to assume that the channel states are binary-valued and chosen by an omniscient adversary. Thus, the adversarial model discussed in Section \ref{online} is a special case of this model with the window-size $w=0$. 
We now propose the following policy which exploits the $w$-step look-ahead information:\\
\underline{Receding Horizon Control} (\textsf{RHC:}) The UE scheduled at each time $t$ is chosen by minimizing the total cost for the next $w$ time-steps. Hence, the scheduling decision at time $t$ is obtained by solving the DP \eqref{opt_dp} with the boundary condition $C^*_{t+w+1}(\bm{h})=0, \forall \bm h$. \\
The \textsf{RHC} policy was considered in \cite{geo_load} in the context of load-balancing in data centers. It was shown that the \textsf{RHC} policy has a competitive ratio of $1+O(\frac{1}{w})$- approaching $1$ as the prediction window size $w$ is increased. Since the result of \cite{geo_load} is not directly applicable to our problem, we examine the gain for AoI due to channel prediction capabilities via numerical simulations in the next section. Unsurprisingly, \textsf{RHC} reduces to the \textsf{MA} policy when the prediction window $w=0.$

\section{Numerical Simulations}\label{simulation}
\begin{figure}
\hspace{-10pt}
\begin{overpic}[width=0.5\textwidth]{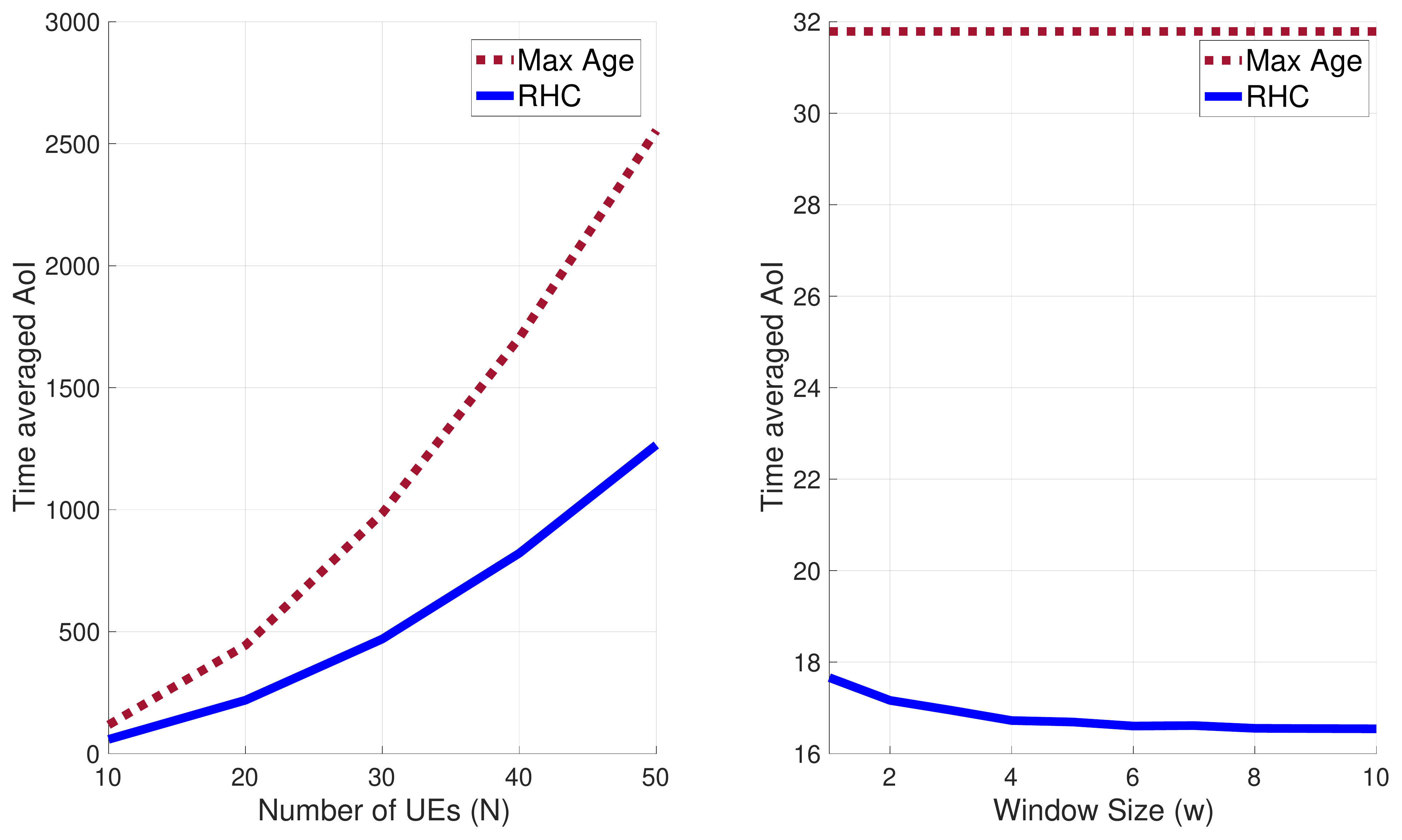}
\put(26,-2){\footnotesize{$(a)$}}
\put(79,-2){\footnotesize{$(b)$}}

\end{overpic}
\caption{Performance comparison between the \textsf{MA} and \textsf{RHC} scheduling policies in a single BS. Figure 1 (a) shows the reduction in the average AoI with as few as $w=3$ slots channel estimations. Figure 1(b) shows the reduction in AoI achieved with $N=5$ UEs as the prediction window $w$ is increased.}
\label{comp_fig}
\end{figure}

In this Section, we perform numerical simulations to compare the performance of the \textsf{RHC} and \textsf{MA} policies in the adversarial setting. Figure \ref{comp_fig} shows the variation of time-averaged AoI with different number of UEs for $T=500$. 
A Monte-Carlo simulation with $k=50$ iterations was performed with randomly generated channels, and we plotted the worst-case AoI in Figure \ref{comp_fig}(a). For each of these iterations, at every time step, the number of \textsf{Good} Channels is selected uniformly at random between $1$ and $N-1$. From the plots, we see that \textsf{RHC} outperforms \textsf{MA} by a large margin even with just a small prediction window of $w=3$. 

 
Figure \ref{comp_fig}(b) shows the variation of the AoI with the window size $(w)$ for the \textsf{RHC} policy. The number of UEs is $N=5$ and the simulation is performed for $T=500$ slots. The window-size is varied from $1$ to $10$. Each simulation is repeated for $50$ times and we plotted the maximum AoI value at the end of these iterations. We see that increasing the prediction window does not significantly decrease the average AoI.

\section{Conclusion and Future Work} \label{conclusion}
This paper investigates the fundamental limits of Age-of-Information in stationary and non-stationary environments from an online scheduling point-of-view. In the stochastic setting, a $2$-optimal scheduling policy has been proposed for mobile UEs. For the non-stationary regime, a new adversarial channel model has been introduced. Upper and lower bounds for the competitive ratio have been derived for the adversarial model. As an immediate extension of this work, the effect of mobility in the non-stationary environment may be considered. The gap between the upper and lower bounds of the competitive ratio may be tightened. Also, it will be interesting to obtain the competitive ratio for $w$-step lookahead policies as a function of the prediction-window $w$. 

\bibliographystyle{IEEEtran}
\clearpage
\bibliography{bibmobility}
\clearpage 
\section{Appendix} \label{appendix}
\subsection{Proof of Theorem \ref{lb}} \label{lb_proof}

\begin{IEEEproof}
In the proof below, we first follow a sample-path-based argument to obtain an almost sure lower bound to AoI. Finally, we use Fatou's lemma \cite{williams1991probability} to convert the almost sure bound to a bound in expected AoI, as defined in Eqn.\ \eqref{objective}. \\
\begin{figure}
\centering
\begin{overpic}[width=0.35\textwidth]{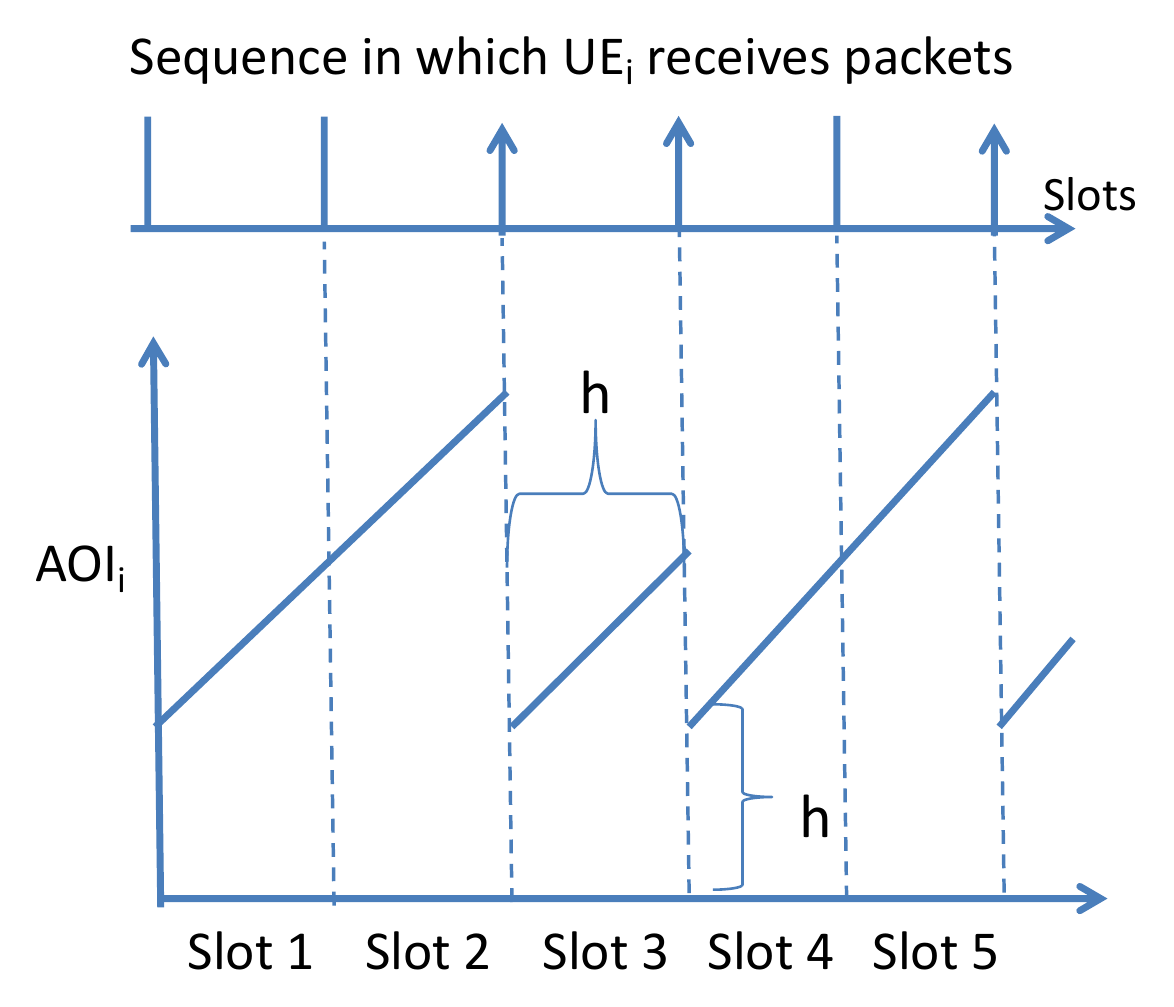}
\end{overpic}
\caption{Time-evolution of the Age-of-Information of a UE}
\label{AoI_fig}
\end{figure}
Consider a sample path under the action of any arbitrary scheduling policy $\pi$ up to time $T$. See Figure \ref{AoI_fig}. For $\textrm{UE}_i$, let the r.v. $N_i(T)$ denote the number of packets received up to time $T$, the r.v. $T_{ij}$ denote the time interval between receiving the $(j-1)$\textsuperscript{th} packet and the $j$\textsuperscript{th} packet, and the r.v. $D_i$ denote the time interval between receiving the last ($N_i(T)$\textsuperscript{th}) packet and the time-horizon $T$. Hence, we have 
\begin{eqnarray} \label{sum_val}
	T= \sum_{j=1}^{N_i(T)} T_{ij} + D_i.
\end{eqnarray}

Since the AoI of any $\textrm{UE}$ increases in step of one at each slot until a new packet is received (and then it drops to one again), the average AoI up to time $T$ may be lower bounded as:
\begin{eqnarray} \label{AoI_lb_der}
	\overline{\textsf{AoI}_T}&\equiv&\frac{1}{NT}\sum_{i=1}^{N} \sum_{t=1}^{T} h_i(t) \nonumber \\
	 &= & \frac{1}{NT}\sum_{i=1}^{N}\bigg(\sum_{j=1}^{N_i(T)} \frac{1}{2}T_{ij}(T_{ij}+1)+ \frac{1}{2}D_i(D_i+1)\bigg) \nonumber \\
	&\stackrel{(a)}{=}&\frac{1}{2NT}\sum_{i=1}^{N}\bigg(N_i(T) \big(\frac{1}{N_i(T)} \sum_{j=1}^{N_i(T)}T_{ij}^2 \big)+D_i^2\bigg)+ \frac{1}{2}\nonumber\\
	&\stackrel{(b)}{\geq} & \frac{1}{2NT}\sum_{i=1}^{N}\bigg( N_i(T)\bar{T_i}^2+D_i^2\bigg)+ \frac{1}{2},
\end{eqnarray}
where in (a) we have used Eqn.\ \eqref{sum_val}, and in (b) 
we have defined $\bar{T}_i= \frac{1}{N_i(T)} \sum_{j=1}^{N_i(T)} T_{ij}$ and used Jensen's inequality afterwards. Rearranging the Eqn. \eqref{sum_val}, we can express the random variable $\bar{T}_i$ as: 
\begin{eqnarray*}
	\bar{T}_i= \frac{T-D_i}{N_i(T)}.
\end{eqnarray*}

\begin{figure}
\centering
\begin{overpic}[width=0.35\textwidth]{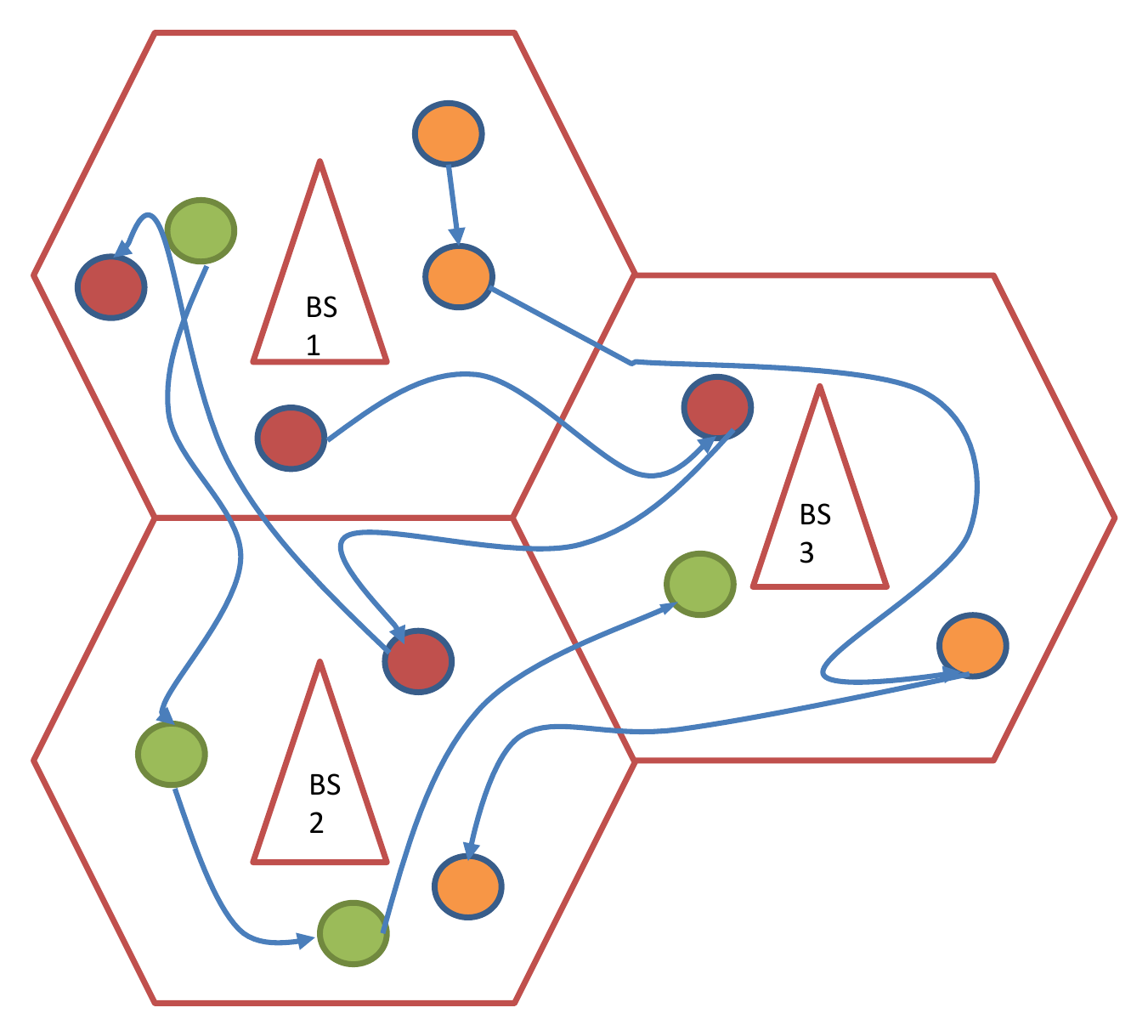}
\end{overpic}
\caption{Movement of $N=3$ UEs in an area with $M=3$ cells }
\label{AoI_mobility_fig}
\end{figure}
With this substitution, the term within the bracket in Equation \eqref{AoI_lb_der} evaluates to 
\begin{eqnarray} \label{AoI_lb_der2}
	 N_i(T)\bar{T}_i^2+ D_i^2 = \frac{(T-D_i)^2}{N_i(T)} + D_i^2 \geq \frac{T^2}{N_i(T)+1},
\end{eqnarray}
where the last inequality is obtained by minimizing the resulting expression by viewing it as a quadratic in the variable $D_i$. \\
Hence, from Eqns.\ \eqref{AoI_lb_der} and \eqref{AoI_lb_der2}, we obtain the following lower bound to the average AoI under the action of any admissible scheduling policy: 
\begin{eqnarray} \label{val}
	\overline{\textsf{AoI}_T} \geq \frac{T}{2N} \sum_{i=1}^{N} \frac{1}{N_i(T)+1} + \frac{1}{2}.
\end{eqnarray}
Next, we analyze the resource constraints of the system to further lower bound the RHS of the inequality \eqref{val}. Let the r.v. $A_i(T)$ denote the total number of transmission attempts made to $\textrm{UE}_i$ by all BSs up to time $T$. Also, let the r.v. $g_j(T)$ denote the fraction of time that $\textrm{BS}_j$ contained \emph{at least one UE} in its coverage area. Since, a BS can attempt a downlink transmission only when there is at least one UE in its coverage area, the total number of transmission attempts to all UEs by the BSs is upper bounded by the following \emph{global balance condition}:
%
%
\begin{eqnarray} \label{attmpt_constr}
\sum_{i=1}^{N} A_i(T) \leq T \sum_{j=1}^{M} g_j(T)\equiv T g(T), 	
\end{eqnarray}
where $g(T) \equiv \sum_j g_j(T)$. Plugging in Eqn.\ \eqref{attmpt_constr}, we can further lower bound the inequality \eqref{val} as:
 \begin{eqnarray*}
\overline{\textsf{AoI}_T} \geq \frac{1}{2Ng(T)} \big(\sum_{i=1}^{N} A_i(T)\big)\big(\sum_{i=1}^{N} \frac{1}{N_i(T)+1}\big) + \frac{1}{2}.
\end{eqnarray*}
An application of the Cauchy-Schwartz inequality on the RHS yields:
\begin{eqnarray} \label{AoI_lb_2}
\overline{\textsf{AoI}_T} \geq\frac{1}{2N g(T)}\bigg(\sum_{i=1}^{N} \sqrt{\frac{A_i(T)}{N_i(T)+1}}\bigg)^2 + \frac{1}{2}.
\end{eqnarray}
 Note that, $\textrm{UE}_i$ successfully received  $N_i(T)$ packets out of a total of $A_i(T)$ packet transmission-attempts made by the BSs via the erasure channel with success probability $p_i$. Without any loss of generality, we may fix our attention on those scheduling policies only for which $\lim_{T \to \infty} A_i(T)= \infty, \forall i$. Otherwise, at least one of the UEs receive a finite number of packets, resulting in infinite average AoI. Hence, using the Strong law of large numbers \cite{williams1991probability}, we obtain:  
 \begin{eqnarray} \label{SLLN2}
 \lim_{T \to \infty} \frac{N_i(T)}{A_i(T)} = p_i, ~~~\forall i \hspace{5pt} \textrm{w.p.} ~ 1. 	
 \end{eqnarray}
Moreover, using the ergodicity property of the UE mobility, we conclude that almost surely: 
\begin{eqnarray*}
\lim_{T \to \infty} g_j(T) = \mathbb{P}_{\bm{\psi}}\big( \textrm{BS}_j \textrm{ contains at least one UE}\big),	
\end{eqnarray*}
where we recall that $\bm{\psi}$ denotes the stationary cell occupancy distribution defined earlier. Thus, we have almost surely 
\begin{eqnarray} \label{g_lim}
\lim_{T \to \infty} g(T) &=& \lim_{T\to \infty} \sum_j g_j(T)\nonumber\\
&=&  \sum_{j=1}^{M} \mathbb{P}_{\bm \psi} \big( \textrm{BS}_j \textrm{ contains at least one UE}\big) \nonumber \\
&\equiv& g(\bm{\psi}), 	
\end{eqnarray}
where the function $g(\bm{\psi})$ denotes the expected number of non-empty cells where the expectation is evaluated w.r.t. the stationary occupancy distribution $\bm \psi$. Hence, putting equations \eqref{SLLN2} and \eqref{g_lim} together with the lower bound in \eqref{AoI_lb_2}, we have almost surely:
\begin{eqnarray} \label{AoI_LB2}
\liminf_{T \to \infty} \overline{\textsf{AoI}_T} \geq 
\frac{1}{2N g(\bm{\psi})}	\bigg(\sum_i \sqrt{\frac{1}{p_i}}\bigg)^2 + \frac{1}{2}.
\end{eqnarray}
Finally, 
\begin{eqnarray*}
\textsf{AoI}^* &\geq& \liminf_{T \to \infty} \mathbb{E}(\textsf{AoI}_T) \\
&\stackrel{(a)}{\geq}& \mathbb{E}(\liminf_{T \to \infty} \textsf{AoI}_T) \\
&\geq& 	\frac{1}{2N g(\bm{\psi})}	\bigg(\sum_i \sqrt{\frac{1}{p_i}}\bigg)^2 + \frac{1}{2},
\end{eqnarray*} 
where the inequality (a) follows from Fatou's lemma. This concludes the proof of Theorem \ref{lb}. Note that the proof continues to hold even when the mobility of the UEs are not independent of each other. 
\end{IEEEproof}

\subsection{Derivation of the bounds in Eqn.\ \eqref{g_unif}}  \label{g_unif_proof}
For $M \geq 2$, we have the following bounds: 
\begin{eqnarray} \label{g_ineq1}
 e^{-\frac{\beta}{M}} \stackrel{(a)}{\leq} (1-\frac{1}{M}) \stackrel{(b)}{\leq} e^{-\frac{1}{M}},  
\end{eqnarray}
where $\beta \equiv \log(4) \leq 1.387.$ 
The inequality (b) is standard. To prove the inequality (a), consider the concave function 
\[f(x) = 1-x-e^{-\beta x}, 0\leq x \leq \frac{1}{2},\]
 for some $\beta > 0$. Since a concave function of a real variable defined on an interval attains its minima at one of the end points of the closed interval, and since $f(0)=0$, we have $f(x) \geq 0, \forall x \in [0, \frac{1}{2}],$ if $f(1/2) \geq 0$, i.e., $ e^{\beta /2} \geq 2$, i.e., $\beta \geq \ln(4)$. Thus, the inequality (a) holds for $M\geq 2$ with $\beta = \ln(4)$. The inequality \eqref{g_ineq1} directly leads to the bounds in Eqn.\ \eqref{g_unif}.
 
 \subsection{Proof of Theorem \ref{achievability_thm}}  \label{achievability_thm_proof}

\begin{IEEEproof}
 Let the scheduling decisions at slot $t$ be denoted by the binary control vector $\bm{\mu}(t) \in \{0,1\}^N$, where $\mu_i(t)=1$ if and only if the following two conditions hold simultaneously: (1) $C_i(t)=j$, \emph{i.e.,} $\textrm{UE}_i$ is within the coverage area of $\textrm{BS}_j$ at slot $t$, for some $1 \leq j \leq M$, and (2) $\textrm{BS}_j$ schedules a transmission  to $\textrm{UE}_i$ at time $t$ \footnote{Recall that the random variable $C_i(t)$ denotes the index of the BS $\textrm{UE}_i$ is associated with at time $t$.}. Since a BS can schedule only one transmission per slot to a UE in its coverage area, the control vector must satisfy the following constraint: 
\begin{eqnarray*}
\sum_{i: C_i(t)=j}\mu_i(t) \leq 1, ~~ \forall j, t. 	
\end{eqnarray*}
 
%
For performance analysis, we consider the following Lyapunov function, which is linear in the ages of the UEs: 
\begin{eqnarray} \label{lyap_linear}
	L(\bm{h}(t))= \sum_{i=1}^N \frac{h_i(t)}{\sqrt{p_i}}.
\end{eqnarray}
The conditional transition probabilities for the age of $\textrm{UE}_i$ may be written as follows:
\begin{eqnarray*}
\mathbb{P}\big(h_i(t+1)=1|\bm{h}(t), \bm{\mu}(t), \bm{C}(t)\big) &=& \mu_i(t)p_i\\
\mathbb{P}\big(h_i(t+1)=h_i(t)+1|\bm{h}(t), \bm{\mu}(t), \bm{C}(t)\big) &=& 1 - \mu_i(t)p_i,
\end{eqnarray*}
where the first equation corresponds to the event when $\textrm{UE}_i$ was scheduled and the packet transmission was successful, and the second equation corresponds to its complement event. 
Hence, for each UE $i$, we can compute :
\begin{eqnarray} \label{one_step_expectation}
	\mathbb{E}\big(h_i(t+1)|\bm{h}(t), \bm{\mu}(t), \bm{C}(t)\big)= h_i(t) -\mu_i(t)p_i h_i(t)+1.
\end{eqnarray}
 From the equation above, we can evaluate the one-step conditional drift as:
 \begin{eqnarray} \label{drift_ineq_1}
 && \mathbb{E}\big(L(\bm{h}(t+1))-L(\bm{h}(t)) | \bm{h}(t), \bm{\mu}(t), \bm{C}(t)\big) \nonumber \\
 &=& -\sum_{i=1}^N \mu_i(t) \sqrt{p_i}h_i(t) + \sum_{i=1}^N \frac{1}{\sqrt{p_i}}. 	
 \end{eqnarray}

Finally, consider the drift minimizing policy \textsf{Multi-Cell MW}  (MMW), under which, each Base Station $\textsf{BS}_{j}$ schedules a user $\textrm{UE}_i$ having the highest weight $\sqrt{p_i}h_i(t)$ in its cell. For the purpose of the proof, we now define a stationary randomized scheduling policy $\textsf{RAND}$, under which every BS randomly schedules a UE in its cell with probability $ \mu^{\textrm{RAND}}_i(t) \propto 1/\sqrt{p_i}$ \footnote{We use the usual convention that summation over an empty set is zero.}. Comparing \textsf{MMW} with \textsf{RAND}, we have:
\begin{eqnarray*}
	\mathbb{E}\bigg(\sum_{i=1}^N \mu^{\textsf{MMW}}_i(t) \sqrt{p_i}h_i(t)| \bm{h}(t), \bm{\mu}(t), \bm{C}(t)\bigg)\\
	\geq \sum_{j=1}^M \frac{\sum_{i: C_i(t)=j} h_i(t)}{\sum_{i: C_i(t)=j} \frac{1}{\sqrt{p_i}}}. 
\end{eqnarray*}
Thus, we have the following upper-bound of the drift \eqref{drift_ineq_1} under the \textsf{MMW} policy:
\begin{framed}
\begin{eqnarray*} 
	\mathbb{E}^{\textsf{MMW}}\big(L(\bm{h}(t+1))-L(\bm{h}(t)) | \bm{h}(t), \bm{C}(t)\big) \\
	\leq -\sum_{j=1}^M \frac{\sum_{i: C_i(t)=j} h_i(t)}{\sum_{i: C_i(t)=j} \frac{1}{\sqrt{p_i}}}+ \sum_{i=1}^N \frac{1}{\sqrt{p_i}}. 
\end{eqnarray*}
\end{framed}
Taking expectation of the above drift-inequality w.r.t. the random cell-occupancy vector $\bm{C}(t)$, we have 
\begin{eqnarray} \label{drift_mob}
	\mathbb{E}^{\textsf{MMW}}\big(L(\bm{h}(t+1))-L(\bm{h}(t))|\bm{h}(t)\big) \nonumber \\
	\leq -\sum_{j=1}^M \mathbb{E}(Z_j(t)|\bm{h}(t)) + \sum_{i=1}^N \frac{1}{\sqrt{p_i}},
\end{eqnarray}
where $Z_j(t) \equiv \frac{\sum_{i: C_i(t)=j} h_i(t)}{\sum_{i: C_i(t)=j} \frac{1}{\sqrt{p_i}}}.$ Our next task is to evaluate this expectation. Note that, we can alternatively express the random variable $\sum_{j=1}^M Z_j(t)$ as 
\begin{eqnarray*}
\sum_{j=1}^M Z_j(t) = \sum_{i=1}^N h_i(t) Y_i(t),	
\end{eqnarray*}
where $Y_i(t) = \big(\frac{1}{\sqrt{p_i}}+\sum_{k\neq i} \frac{1}{\sqrt{p_k}}\mathds{1}(C_i(t)=C_k(t))\big)^{-1}. $ 

%
%
%
%
%
%
%
We can evaluate this expectation exactly for the i.i.d. uniform mobility model. Recall that $\bm{C}(t) \perp \bm{h}(t)$. Hence,
\begin{eqnarray} \label{yi}
\mathbb{E}(Y_i(t)) = \sum_{n=0}^{N-1}\sum_{S: i\notin S, |S|=n} \bigg(\frac{1}{\sqrt{p_i}}+\sum_{k\in S} \frac{1}{\sqrt{p_k}}\bigg)^{-1} \times \nonumber \\
\frac{1}{M^n}\bigg(1-\frac{1}{M}\bigg)^{N-n-1}.
\end{eqnarray}
In the special case when all UEs are identical, \emph{i.e.,} $p_i=p, \forall i$, the summation \eqref{yi} has a closed-form expression. Clearly, for all $ 0 \leq n \leq N-1$, we have:
\begin{eqnarray*}
Y_i(t)= \frac{\sqrt{p}}{n+1},~~~ \textrm{w.p.}~ \binom{N-1}{n} \frac{1}{M^n}\bigg(1-\frac{1}{M}\bigg)^{N-n-1}.
\end{eqnarray*}
To evaluate the expectation of $Y_i(t)$, we integrate the binomial expansion of $(1+x)^{N-1}$ in the range $[0,\beta]$ to obtain the identity:

\begin{eqnarray*}
\frac{1}{N}\bigg( (1+\beta)^N-1 \bigg) = \beta \sum_{n=0}^{N-1} \frac{1}{n+1}\binom{N-1}{n} \beta^n. 	
\end{eqnarray*}
Substituting $\beta = \frac{1}{M-1}$ in the above, we obtain 
\begin{eqnarray}\label{iid_yi}
\mathbb{E}(Y_i(t))= \sqrt{p}\frac{M}{N}\bigg(1-\big(1-\frac{1}{M}\big)^{N} \bigg) \equiv Y^*(\textrm{say}). 	
\end{eqnarray}
From Eqn. \eqref{drift_mob} and \eqref{iid_yi}, we have 
\begin{eqnarray*}
\mathbb{E}^{\textsf{MMW}}\big(L(\bm{h}(t+1))-L(\bm{h}(t))|\bm{h}(t)\big) \leq -Y^*\sum_i h_i(t) + \frac{N}{\sqrt{p}}.	
\end{eqnarray*}
Taking expectation of both sides, we have 
\begin{eqnarray*}
	\mathbb{E}^{\textsf{MMW}}\big(L(\bm{h}(t+1))-L(\bm{h}(t))\big) \leq -Y^*\sum_i \mathbb{E}h_i(t) + \frac{N}{\sqrt{p}}.
\end{eqnarray*}
Summing up the above inequalities and averaging w.r.t. $T$ slots, we obtain

\begin{eqnarray}\label{ub_pf_1}
	\textsf{AoI}^{\textsf{MMW}}&=&\limsup_{T \to \infty} \frac{1}{NT}\sum_{t=1}^{T}\sum_i \mathbb{E}h_i(t) \nonumber \\
	&\leq& \frac{N}{Y^*\sqrt{p}}= \frac{N}{Mp\bigg( 1- (1-\frac{1}{M})^N\bigg)}.
\end{eqnarray}
On the other hand, the lower bound from Theorem \ref{lb}, specialized to this case, yields:
\begin{eqnarray} \label{lb_pf_1}
\textsf{AoI}^* \geq \frac{N}{2Mp\bigg( 1- (1-\frac{1}{M})^N\bigg)}.	
\end{eqnarray}
Eqns.\ \eqref{ub_pf_1} and \eqref{lb_pf_1}, we have 
\[\textsf{AoI}^{\textsf{MMW}} \leq 2\textsf{AoI}^*. \]
The above inequality shows that the policy $\textsf{MMW}$ is $2-$optimal in the case of statistically identical UEs with uniform mobility.

%
%
%
\end{IEEEproof}
\begin{figure}
\centering
\begin{overpic}[width=0.5\textwidth]{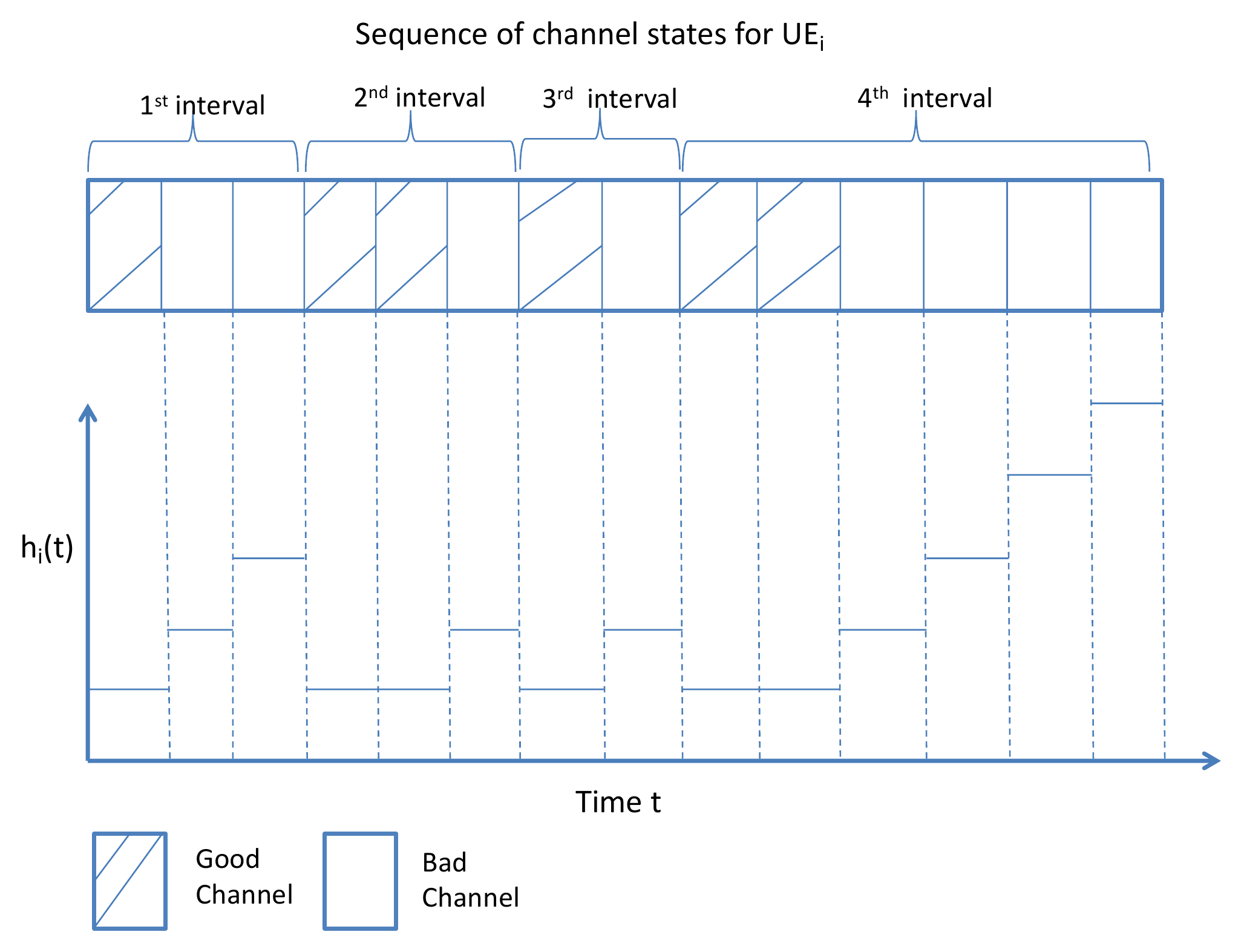}
\end{overpic}
\put(-221,184){\footnotesize{$\Delta_1$}}
\put(-74,184){\footnotesize{$\Delta_4$}}

\caption{\small {Illustrating the \emph{intervals} for $\textrm{UE}_i$}}
\label{intervals_fig}
\end{figure}

\subsection{Proof of Theorem \ref{comp_ratio_ub}} \label{comp_ratio_ub_proof}
\begin{IEEEproof}
	Let us assume that the \textsf{MA} policy had $K \geq 0$ successful transmissions during the entire time-horizon of length $T$. We divide the time horizon into $K$ successive \emph{intervals}, defined naturally as follows. Let $T_i$ be the time index at which the \textsf{MA} policy had its $i$\textsuperscript{th} successful transmission, $0\leq i \leq K$, and $T_{K+1}=T$. Let $\Delta_i \equiv T_i- T_{i-1}$ denote the length of the $i$\textsuperscript{th} interval between the $i$\textsuperscript{th} and $i-1$ \textsuperscript{th} successful transmissions of the \textsf{MA} policy. For notational consistency, we define $T_0 \equiv 0, \Delta_0 \equiv 0.$  See Figure \ref{intervals_fig}.
%
%
	We start our analysis with two simple observations - first, whenever a successful transmission is made by the \textsf{MA} policy, the optimal policy \textsf{OPT} also transmits at that slot successfully. Second, the \textsf{MA} policy is a \emph{persistent round robin} policy, which keeps on scheduling a user (having the highest age) until the transmission is successful. In the immediately following time slot, the \textsf{MA} policy switches to the other user and continues the round-robin scheduling cycle. See Figure \ref{OPT_MW_fig} for a typical run. 
	\begin{figure}
\centering
\begin{overpic}[width=0.5\textwidth]{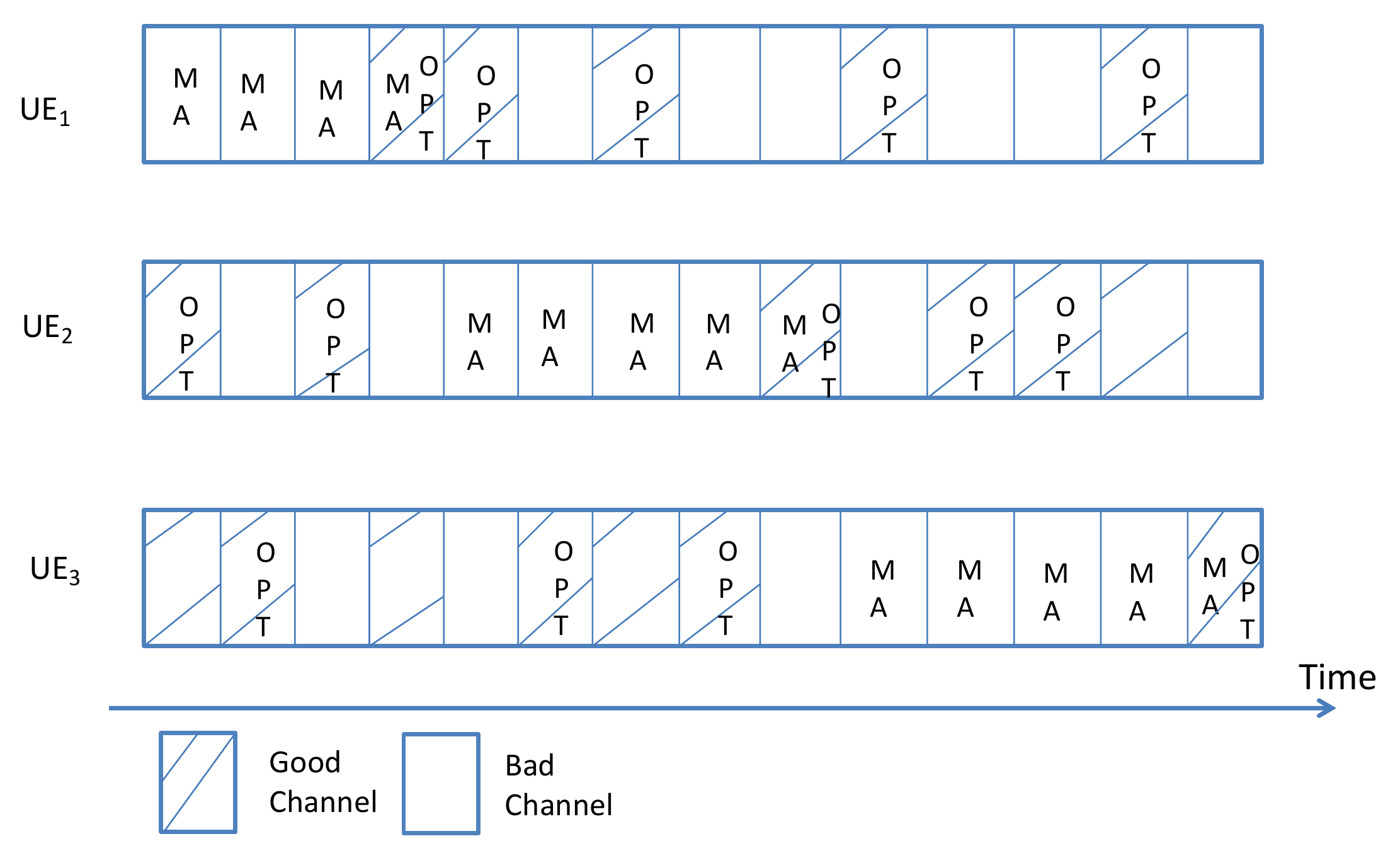}
\end{overpic}

\caption{Illustrating the scheduling decisions of \textsf{MA} and \textsf{OPT} with $N=3$ UEs. User which is scheduled by the \textsf{MA} policy at each slot is denoted by MA and the user which is scheduled by the offline Optimal policy \textsf{OPT} at each slot is denoted by OPT and the user which is scheduled by both \textsf{MA} and \textsf{OPT} at the same instant is denoted by MA as well as OPT. The figure shows that the \textsf{MA} policy sticks to one user till it gets served and then it switches over to another user in a round-robin fashion. This figure also shows that how the optimal algorithm takes advantage of the known channel states.}
\label{OPT_MW_fig}
\end{figure}
	
	Hence, under the \textsf{MA} policy, the states of the users (in sorted order) at the beginning of the $i$\textsuperscript{th} interval is \[\{1, 1+\Delta_{i-1}, 1 + \Delta_{i-1}+ \Delta_{i-2}, \ldots, 1+ \sum_{j=1}^{N-1} \Delta_{i-j}.\}\] 
	Since the \textsf{MA} policy continues scheduling the UE having the highest age, at the end of the $k$\textsuperscript{th} slot of the $i$\textsuperscript{th} interval, the ages of the UEs (in sorted order) are given by: 
	\[\{k, k+\Delta_{i-1}, k + \Delta_{i-1}+ \Delta_{i-2}, \ldots, k+ \sum_{j=1}^{N-1} \Delta_{i-j}, ~~ 1\leq k \leq \Delta_i.\}\]
 Hence, the cost $C_i^{\textsf{MA}}$ incurred by the \textsf{MA} policy during the $i$\textsuperscript{th} interval is computed as: 
	\begin{eqnarray} \label{CMA}
	C_i^{\textsf{MA}}&=& \sum_{k=1}^{\Delta_i} k + \sum_{k=1}^{\Delta_i}\sum_{m=1}^{N-1}\bigg(k+\big(\sum_{j=1}^{m} \Delta_{i-j}\big)\bigg)\nonumber \\
	&=&N \sum_{k=1}^{\Delta_i}k + \Delta_i \sum_{j=1}^{N-1}(N-j)\Delta_{i-j} \nonumber\\
	&\leq & N \bigg( \frac{\Delta_i(\Delta_i+1)}{2} + \sum_{j=1}^{N-1} \Delta_i \Delta_{i-j} \bigg)\\
	&\leq & \frac{N}{2}\bigg( N \Delta_i^2 + \Delta_i+ \sum_{j=1}^{N-1} \Delta_{i-j}^2\bigg)
	\end{eqnarray}
	where in the last step, we have used the AM-GM inequality to conclude $ \Delta_i \Delta_{i-j} \leq \frac{1}{2}\big(\Delta_i^2 + \Delta_{i-j}^2\big), 1\leq j \leq N-1.$\\
	Hence, the total AoI cost incurred  by the \textsf{MA} scheduling policy over the entire time horizon is upper bounded as: 
	\begin{eqnarray*}
	\textrm{AoI}^{\textsf{MA}}(T)&=& \sum_{i=1}^{K}C_i^{\textsf{MA}} \\
	&\leq & \frac{N}{2}\sum_{i=1}^{K}\bigg( N \Delta_i^2 + \Delta_i+ \sum_{j=1}^{N-1} \Delta_{i-j}^2\bigg)\\
	&\leq & \frac{N}{2}\sum_{i=1}^{K} \bigg(2N \Delta_i^2 + \Delta_i\bigg).
	\end{eqnarray*}

On the other hand, the cost incurred by \textsf{OPT} during the $i$\textsuperscript{th} interval is lower bounded as: 
	\begin{eqnarray} \label{COPT}
	C_i^{\textsf{OPT}}&\geq & (N-1)\sum_{k=1}^{\Delta_i}1 + \sum_{k=1}^{\Delta_i} (1+k). \nonumber\\
	&\geq & \frac{1}{2} \Delta_i^2 + N\Delta_i, 
	\end{eqnarray}
	where we have separately lower bounded the cost incurred by the UE being scheduled by \textsf{MA} (which was consistently seeing \textsf{Bad} channels) and the other UEs. 
Finally, the cost of the entire horizon may be obtained by summing up the cost incurred in the constituent intervals. 
	Hence, noting that $\Delta_0=0$, from Eqns.\ \eqref{CMA} and \eqref{COPT}, the competitive ratio $\eta^{\textsf{MA}}$ of the \textsf{MA} policy may be upper bounded as follows:
	\begin{eqnarray*}
	\eta^{\textsf{MA}} &=& \frac{\sum_{i=1}^K C_i^{\textsf{MA}}}{\sum_{i=1}^K C_i^{\textsf{OPT}}}	\\
	&\stackrel{(a)}{\leq}& \frac{\frac{N}{2}\sum_{i=1}^{K} \bigg(2N \Delta_i^2 + \Delta_i\bigg)}{\sum_{i=1}^K \big(\frac{1}{2}\Delta_i^2 + N\Delta_i\big)} \\
	& \leq & 2N^2.
	\end{eqnarray*}

\end{IEEEproof}
\subsection{Proof of Theorem \ref{comp_ratio_lb}} \label{comp_ratio_lb_proof}
\begin{IEEEproof}

To apply Yao's principle, we need to compute the expectations appearing in the numerator and the denominator of Eqn.\ \eqref{Yao_lb}.  
\subsubsection{Upper bound to \textsf{OPT}'s expected cost}
Let the random variable $C_i(T)$ denote the total AoI-cost incurred by the $i$\textsuperscript{th} UE up to time $T$. In other words, 
\begin{eqnarray*}
	C_i(T) = \sum_{t=1}^{T} h_i(t).
\end{eqnarray*}
Hence, the limiting time-averaged total expected cost incurred by \textsf{OPT} may be expressed as 
\begin{eqnarray} \label{opt_ub_1}
\bar{\mathcal{C}}(\textsf{OPT}) \equiv	\lim_{T \to \infty} \frac{1}{T} \sum_{i=1}^{N} \mathbb{E}\big(C_i(T)\big) = \sum_{i=1}^{N} \lim_{T \to \infty} \frac{\mathbb{E}(C_i(T))}{T},  
\end{eqnarray}
In the following, we will show that all of the above limits exist with the assumed choice of the underlying probability space.
We now use the Renewal Reward Theorem \cite{gallager2012discrete} in order to evaluate the RHS of Eqn.\ \eqref{opt_ub_1}. Since, under the assumed channel state distribution $\bm p$, only one channel is in \textsf{Good} state, the optimal policy \textsf{OPT} is easy to characterize - at any slot, \textsf{OPT} schedules the user having \textsf{Good} channel.
 Under this probability space, it can be verified that, for each user $i$, the sequence of random variables $\{h_i(t)\}_{t\geq 1}$ constitute a renewal process, with the commencement of scheduling of the $i$\textsuperscript{th} user constituting renewal instants. A generic renewal interval of length $\tau$  for the $i$\textsuperscript{th} user consists of two parts - (1) a consecutive sequence of \textsf{Good} channels of length $\tau_\textsf{G}$, and (2) a consecutive sequence of \textsf{Bad} channels of length $\tau_{\textsf{B}}$.  The AoI cost $c_i(\tau)$ incurred by the user $i$ in any generic renewal cycle may be written as the sum of the costs incurred in two parts:  
 \begin{eqnarray*}
c_i(\tau)&=& c_i(\tau_{\textsf{G}})+ c_i(\tau_{\textsf{B}})\\
&=& \sum_{t=1}^{\tau_{\textsf{G}}}1 +  \sum_{t=1}^{\tau_{\textsf{B}}}(1+t)\\
&=& \tau_{\textsf{G}} + \frac{3}{2}\tau_{\textsf{B}} + \frac{1}{2}\tau_{\textsf{B}}^2. 
\end{eqnarray*}
Let $q\equiv \frac{1}{N}$ be the probability that that the channel is \textsf{Good} for the $i$\textsuperscript{th} user at any slot. Hence, from our construction, the random variables $\tau_{\textsf{G}}$ and $\tau_{\textsf{B}}$ follows a Geometric distribution having the following p.m.f. 
\begin{eqnarray*}
\mathbb{P}(\tau_{\textsf{G}}=k) &=& q^{k-1} (1-q), ~~ k \geq 1. \\	
\mathbb{P}(\tau_{\textsf{B}}=k) &=& q(1-q)^{k-1}, ~~ k \geq 1. 
\end{eqnarray*}
Hence, the expected cost incurred by the $i$\textsuperscript{th} user at any renewal cycle is given by 
\begin{eqnarray}\label{cycle_cost}
	\mathbb{E}(c_i(\tau))= \frac{1}{1-q}+ \frac{3}{2q}+ \frac{2-q}{2q^2}= \frac{1}{q^2(1-q)}.
\end{eqnarray}
Moreover, the expected length of any renewal cycle is given by 
\begin{eqnarray}\label{cycle_length}
\mathbb{E}(\tau)= \mathbb{E}(\tau_{\textsf{G}})+ \mathbb{E}(\tau_{\textsf{B}})= \frac{1}{q(1-q)}.	
\end{eqnarray}

Using Renewal Reward Theorem \cite{gallager2012discrete}, we have 
\begin{eqnarray*}
\lim_{T \to \infty} \frac{\mathbb{E}(C_i(T))}{T}= \frac{\mathbb{E}(c_i(\tau))}{\mathbb{E}(\tau)}=\frac{1}{q}=N, ~~~\forall i.	
\end{eqnarray*}
Hence, from \eqref{opt_ub}, we conclude that the time-averaged total expected cost incurred by \textsf{OPT} is given by 
\begin{eqnarray}\label{opt_ub}
	\bar{\mathcal{C}}(\textsf{OPT})= N^2. 
\end{eqnarray}

\subsubsection{Lower Bound to the AoI for $N$ users}
By directly appealing to the general lower bound in Theorem \eqref{lb}, with $p_i=\frac{1}{N}, ~\forall i$, and $M=1$, we conclude that under the assumed channel state distribution, the time-averaged expected cost for any online scheduling policy $\pi$ is lower bounded as 
\begin{eqnarray} \label{lb_gnl}
\bar{\mathcal{C}}(\pi) = \limsup_{T\to \infty} \frac{1}{T}\sum_{i=1}^{N}\mathbb{E}(C_i(T)) \geq \frac{N^3+N}{2}.	
\end{eqnarray}
We should point out that the lower bound in \eqref{lb_gnl} is not numerically tight. In particular, the following Proposition \ref{improved_LB} shows that, using a more careful analysis, the AoI lower bound for $N=2$ users may be improved to $6$. 
\begin{framed}
\begin{proposition} \label{improved_LB}
In the above set up, for any online policy, the average AoI for $N=2$ users with the probability of successful transmission $p_1=p_2=\frac{1}{2}$ is lower bounded by $6$.\end{proposition}
\end{framed}
For a proof of the above proposition, please refer to Appendix \ref{improved_LB_proof} below.\\
Nevertheless, the achievability result in Theorem \ref{achievability_thm} shows that the bound in Eqn.\ \eqref{lb_gnl} is tight within a factor of $2$. In particular, Eqn.\ \eqref{lb_gnl} has the order optimal dependence on $N$.
Finally, using Yao's minimax principle in conjunction with Eqns. \eqref{opt_ub} and \eqref{lb_gnl}, we conclude that the competitive ratio $\eta(N)$ of any online policy is lower bounded as 
\begin{eqnarray*}
\eta(N) \geq \sup_{T}\frac{C_T(\pi)}{C_T(\textsf{OPT})}	\geq \frac{\bar{\mathcal{C}}(\pi)}{\bar{\mathcal{C}}(\textsf{OPT})} \geq \frac{N}{2} + \frac{1}{2N}. 
\end{eqnarray*}
In the case when $N=2$, using the result of Appendix \ref{improved_LB_proof}, the competitive ratio is lower bounded by 
\[ \eta(2) \geq \frac{6}{2^2}=1.5.\]
\end{IEEEproof}

\subsection{Proof of Proposition \ref{improved_LB}} \label{improved_LB_proof}
\begin{IEEEproof}

Define $\mathcal{F}_{t-1}\equiv \sigma(\vec{h}(k), \vec{\mu}(k), 1\leq k \leq t-1)$ to be the sigma-algebra generated by the r.v.s of age and control vectors observed up to time $t-1$. Since the policy is online, the scheduling decision $\vec{\mu}(t)$ at time $t$ must be measurable in $\mathcal{F}_{t-1}$ for all $t\geq 1$.  Let $H_{\textrm{sum}}(t) \equiv \mathbb{E}^\pi(h_1(t))+ \mathbb{E}^\pi(h_2(t))$ be the expected sum of the ages of the UEs at time $t$. Let $B_t \in \mathcal{F}_t$ be the event for which the $\textrm{UE}_1$ is scheduled under the policy $\pi$. Then, we can write 
\begin{eqnarray} \label{cond_ex1}
&&\mathbb{E}^\pi\big(h_1(t+1)|\mathcal{F}_t) \\
&=& \big(1+\frac{1}{2}h_1(t)\big)\mathds{1}(B_t) + \big(1+h_1(t\big) \mathds{1}(B_t^c) \nonumber \\
 &=& 1+ \frac{1}{2}h_1(t) + \frac{1}{2}h_1(t)\mathds{1}(B_t^c)\nonumber \\
 &\stackrel{(a)}{\geq} & 1+ \frac{1}{2}h_1(t) + \frac{1}{2}\min \{h_1(t), h_2(t)\}\mathds{1}(B_t^c),
\end{eqnarray}
Similarly, we can also write
\begin{eqnarray}\label{cond_ex2}
	\mathbb{E}^\pi\big(h_2(t+1)|\mathcal{F}_t) \geq 1+ \frac{1}{2}h_2(t) + \frac{1}{2}\min\{h_1(t), h_2(t) \}\mathds{1}(B_t).
\end{eqnarray}
Since $\mathds{1}(B_t)+\mathds{1}(B_t^c)=1$, from the equations \eqref{cond_ex1} and \eqref{cond_ex2}, we have 
\begin{eqnarray*}
	&&\mathbb{E}^\pi\big(h_1(t+1)+h_2(t+1)|\mathcal{F}_t) \geq\\
	&& 2+ \frac{1}{2}(h_1(t)+h_2(t))+ \frac{1}{2} \min\{h_1(t), h_2(t)\}.
\end{eqnarray*}
Taking expectations of both sides of the above equation, we get
\begin{eqnarray} \label{key_eqn1}
H_{\textrm{sum}}(t+1) \geq 2 + \frac{1}{2} H_{\textrm{sum}}(t) + \frac{1}{2} \mathbb{E}\bigg(\min\{h_1(t), h_2(t)\}\bigg).	
\end{eqnarray}
Let the random variable $S(t)$ denote the time elapsed since the last successful transmission (by any UE) before time $t$. Clearly, \[ \min \{h_1(t), h_2(t)\} \geq S(t)\] (the above inequality holds with equality for the two user case).
Hence, the above inequality implies 
\[ H_{\textrm{sum}}(t+1) \geq 2 + \frac{1}{2} H_{\textrm{sum}}(t) + \frac{1}{2} \mathbb{E}\big(S(t)\big).\]
Summing up the above inequalities for $t=1,2, \ldots, T$, and dividing both sides by $T$, we obtain 
\begin{eqnarray} \label{cesaro_mean_lt}
	2\frac{H_{\textrm{sum}}(T+1)}{T}+\frac{1}{T}\sum_{t=1}^{T}H_{\textrm{sum}}(t) \geq 4 + \frac{1}{T}\sum_{t=1}^{T}\mathbb{E}(S(t)).
\end{eqnarray}
  It is to be noted that $\{S(t)\}_{t\geq 1}$ is a renewal process with the time-stamp of successful transmissions constituting the renewal instants. Let the random variable $\tau$ denote the length of any generic renewal cycle. Hence, using the renewal reward theorem \cite{gallager2012discrete} \cite{gallager2013stochastic}, it follows that 
\begin{eqnarray*}
	\lim_{T \to \infty} \frac{1}{T}\sum_{t=1}^{T}\mathbb{E}(S(t)).&=& \frac{\mathbb{E}\big(\int_{0}^{\tau}S(t)dt\big)}{\mathbb{E}(\tau)}\\
	&=& \frac{\mathbb{E}(1+2+\ldots+\tau)}{\mathbb{E}(\tau)}\\
	&=& \frac{\mathbb{E}(\tau^2)+\mathbb{E}(\tau)}{2\mathbb{E}(\tau)}\\
	&=& 2,
\end{eqnarray*}

where the last inequality follows from the fact that the
\newpage
 renewal cycle lengths $T$ are distributed geometrically with the parameter $p=1/2$. Thus, the limit of the RHS of Eqn.\ \eqref{cesaro_mean_lt} exists and the limiting value is equal to $6$.
Next, we consider two possible cases. \\
\textbf{Case I: $\liminf_{T\to \infty} \frac{H_{\textrm{sum}}(T+1)}{T}=0$: } In this case, consider a subsequence $\{T_k\}_{k\geq 1}$ along which $\lim_{k\to \infty} \frac{H_{\textrm{sum}}(T_k+1)}{T_k}=0$. 

For this subsequence, we have from Eqn.\ \eqref{cesaro_mean_lt}:
\begin{eqnarray*}
	2\frac{H_{\textrm{sum}}(T_k+1)}{T_k}+\frac{1}{T_k}\sum_{t=1}^{T_k}H_{\textrm{sum}}(t) \geq 4 + \frac{1}{T_k}\sum_{t=1}^{T_k}\mathbb{E}(S(t)).
\end{eqnarray*}
Taking $k \to \infty$, we conclude that 
\begin{eqnarray}\label{ces_lim2}
\limsup_{T\to \infty} \frac{1}{T}\sum_{t=1}^{T}H_{\textrm{sum}}(t) \geq 6.
\end{eqnarray}
\textbf{Case II: $\liminf_{T\to \infty} \frac{H_{\textrm{sum}}(T+1)}{T}=\alpha >0$: }
From the definition of $\liminf$, it follows that there exists a finite $T_0$ such that, for all $T \geq T_0$, we have 
\begin{eqnarray} \label{liminf_eqn}
\frac{H_{\textrm{sum}}(T+1)}{T} \geq \frac{\alpha}{2}. 	
\end{eqnarray}
Thus, for any $T \geq T_0$, we can write 
\begin{eqnarray*}
	\frac{1}{T}\sum_{t=1}^{T} H_{\textrm{sum}}(t) \geq \frac{1}{T}\sum_{t=T_0+1}^{T} H_{\textrm{sum}}(t) \stackrel{(a)}{\geq} \frac{\alpha}{2T}\sum_{t=T_0}^{T-1}t = \Omega(T).
\end{eqnarray*}
Hence, in this case, we have 
\begin{eqnarray*}\label{lim_cost1}
\limsup_{T \to \infty} \frac{1}{T}\sum_{t=1}^{T} H_{\textrm{sum}}(t)  = \infty.	
\end{eqnarray*}
Hence, from Eqns. \eqref{ces_lim2} and \eqref{lim_cost1}, we conclude that, in either case, we have 
\begin{eqnarray} \label{lim_cost2}
	\limsup_{T \to \infty} \frac{1}{T}\sum_{t=1}^{T} H_{\textrm{sum}}(t)  \geq 6. 
\end{eqnarray}
\end{IEEEproof}

\end{document}